\documentclass[12pt,notitlepage]{article}
 \usepackage{color}
\usepackage{amsmath}
\usepackage{amsthm}
\usepackage{epsfig}
\usepackage{latexsym}
\usepackage{pdflscape}
\numberwithin{equation}{section}
\usepackage{graphicx}
\title{\Large \bf  Day of the week effect in
paper submission/acceptance/rejection to/in/by \\ peer review journals. \\Ê 
 II. An ARCH econometric-like modeling  }
\author{ Marcel Ausloos$^{1,2}$, \\ Olgica Nedic$^{3}$, \\   Aleksandar Dekanski$^{4}$, \\ 
Maciej J. Mrowinski$^{5,¬*}$, Piotr Fronczak$^{5,¬**}$, Agata Fronczak$^{5,¬¬*¬*¬*}$  }

 \date{$^1$  School of Management, University of Leicester, University Road, Leicester  LE1 7RH, UK;
 \\$e$-$mail$ $address$: ma683@le.ac.uk \\
$^2$GRAPES\footnote{Group
of Researchers for Applications of Physics in Economy and Sociology}$\;$,
  rue de la Belle Jardiniere 483, \\B-4031, Angleur, Belgium \\$e$-$mail$ $address$:
marcel.ausloos@ulg.ac.be
\\ 
$^3$ Institute for the Application of Nuclear Energy (INEP),
University of Belgrade, Banatska 31b, Belgrade-Zemun, Serbia
   \\$e$-$mail$ $address$: olgica@inep.co.rs \\ 
$^4$ Institute of Chemistry, Technology and Metallurgy,
Department of Electrochemistry, University of Belgrade, 
Njegoseva12, Belgrade, Serbia
 \\$e$-$mail$ $address$: dekanski@ihtm.bg.ac.rs
 \\
 $^{5}$Faculty of Physics, Warsaw University of Technology,  \\Koszykowa 75, PL-00-662, Warsaw, Poland 
 \\ 
 $^{¬*}$ $e$-$mail$ $address$: mrow@if.pw.edu.pl 
 \\ 
 $^{¬*¬*}$  $e$-$mail$ $address$: fronczak@if.pw.edu.pl
 \\
 $^{¬*¬*¬*}$   $e$-$mail$ $address$: agatka@olimp.if.pw.edu.pl }

\begin{document}
 \maketitle
\begin{abstract}
  This paper aims at providing a statistical model for the preferred behavior of authors submitting a paper to a scientific journal. The electronic
submission of (about 600) papers to the Journal of the Serbian Chemical
Society has been recorded for every day from Jan. 01, 2013 till Dec. 31,
2014, together with the acceptance or rejection paper fate. Seasonal effects
and editor roles (through desk rejection and subfield editors) are examined. An
ARCH-like econometric model is derived stressing the main determinants
of the favorite day-of-week process. \end{abstract}  
  
    \vskip0.5cm
\section{Foreword}

Quantitative considerations on human aspects of synchronized behavior or cyclic rhythms are abundant:  e.g., menstruation or heart beat \cite{mcclintock71} - \cite{roehner2007driving};
 specific days of week effects are reported for many  issues: birth rates, judges' decisions, car accidents, thieves activity, hospital admission, mortality rate, or when women are feeling to be least attractive \cite{HumRep} - \cite{women}
 ; for financial markets  \cite{French} - \cite{Cellini_AFE24.14.161} 
 a day-of-the-week effect   is very well known.  Of course, on all such  findings, we are aware that statistical critiques are numerous.

In order to acquire information, intended to monitor (hidden) psychological motivations,  it is of interest  to focus some attention on similar questions outside  the financial and economic sphere. There are many possible cases \cite{roehner2007driving}, but the data must be first realistically obtained, next it should be reliable, and {\it  in fine} has to  suggest more investigations within a broadening concern about human behavior. 

We have been fortunate to get access to such data   about submitted, accepted or  rejected papers  in the peer review  process of     the  Journal of the Serbian Chemical Society   (JSCS). We have examined such data in \cite{TWeffectPhA496.16.197}.  A  behavioral hypothesis about submitting authors was sketched.  However, as mentioned by reviewers and others, there was no  subsequent model  "describing" the  findings. 
Nevertheless, 
  the outlined hypothesis,  based on  some expected behavior(s) of  scientists, taking into account their work environment,  seemed reasonable, suggesting {\it mutatis mutandis} some universal feature for  when    manuscripts  are submitted to a journal \cite{TWeffectPhA496.16.197}, related to  the paper   quality influencing editors and reviewers appreciation, whence leading to acceptance or rejection. 
  
  However,  to propose a behavioral model is not a trivial or common feature when physics rigor is expected. One may mention work in sociophysics \cite{galammoscovici1991,ContucciGhirlandaQuQu2007,MAsocialforcesACS013}, but the pertinence of such models can be debated, because of external factors, as pointed for example  in  \cite{Weinshall-MargelShapard,kulakowskiparameters} and because it is unclear that collective (herding) causes are the primary source of behaviors. The matter is delicate. We understand that models which intend to capture reality through fitting parameters  are much scorned upon in physics realms. However, regression models may also present some interest in order to emphasize significant variables \cite{HumRep}. Here below, we propose such  a model, based on econometrics technique.

\section{Introduction}
 
A peer-review process starts when a paper is submitted to a journal and ends when the paper is accepted or rejected for publication. How    reviewers behave has already been much studied. The intellectual writing process after compiling  measures and their subsequent analysis is also quite studied. However, a major step  occurring when  presenting research results is far less studied.  This is studied here below, - the submission day, together with some practical measure of its consequences: its influence on the acceptance (or rejection) of the manuscript.

Due to electronic submissions nowadays, the submission  is  quasi entirely managed by an author of the paper.    It is logical to admit that the behavior is often  influenced by the action of others, but could also be intrinsic due to societal constraints or habits \cite{postautistic}. Beside such a behavioral aspect, it  seems of practical interest, for editors, reviewers and publishers,  to
 explore the timely behavior of  agents submitting papers  in scientific journals and the editor work flow  \cite{MFFNAeditwork}.  It is usual to find some information on the date of submission of a paper in recent years on the first page of a paper. The date of acceptance is also often announced, but the latter depends on many individual factors, inherent to the editorial peer review and process. What is hardly known is the  number of papers which are submitted,  - on a given day, whatever their later fate. What is quasi unknown is the number  of papers  which are rejected, after having been submitted on a given day. These last two numbers of submissions, whence the information on the day of submission are  \underline{entirely and strictly }an author behavioral measure with respect to his/her research work.  It is not influenced by  reviewers or editors (except maybe for special issues with deadlines). Thus, we stress that even though the acceptance or rejection does not entirely depend on the authors, the submission is in his/her hands only.

 We have been fortunate to get access to   data on submission, acceptance or also rejection  from  the  Journal of the Serbian Chemical Society   (JSCS): about 265  papers  were accepted along  600  submitted papers over  2  years, i.e. 730 days.   The journal contains various sub-sections.  It had  an impact factor = 0.912 in 2012, - before the years 2013 and 2014,  those  in which the peer review  starting days are examined below.  N.B. One might contrast these quantities  with the order of  magnitude known for   another respected  journal, Nature  which recently  reported 3089 of submissions over 10 weeks;  this means  about 310 papers per week. Only  4 articles and 13 letters were published, e.g., in one of  the latest issues. This means a   4\% acceptance rate (or 96\% rejection rate) \cite{TWeffectPhA496.16.197}.   However,  we stress that  Nature and other journals have not released, to our knowledge, any  information  neither on the day of submission nor on the fate of the manuscript according to such a day of submission.  Therefore our JSCS data present  unique features.

  We find that, in the JSCS  case at hands, more papers are submitted on  Wednesday,  but when examining the relative acceptance rate, with respect  to the total number of submitted papers in a given week day,   more papers are accepted for publications if submitted on Tuesday.  There is no information on the day of acceptance or rejection by an editor.  Only the submission and paper fate process are thus considered as the dependent variables. 
  
  To develop a behavioral (agent based-like) model  seems  too audacious. Let it be observed that what is presently examined is not the peer review process {\it per se}, for which several physics-based models have been already outlined, mainly stressing either the editor side \cite{SpringerPlus5.16.903Wang} or the reviewer side \cite{EPJB84.11.707hanel,Scim106.16.695Ð715kovanis}, or both \cite{MFFNAeditwork}. A "weaker"   approach, like a statistical modeling,  can be envisaged, as in econometrics:  an ARCH-like modeling  
  \cite{AFE10.00.482},  \cite{bollerslev1994arch}-\cite{GibbonsHess}, 
   - here emphasizing the \underline{authors preferred day} of submission as well as the \underline{most successful day} of submission for paper acceptance.  We consider that we may take the  same modeling approach   used to describe   an investor behavior   differences (on Monday and Friday with respect to other days of the week), 

  Time series  are shown in Sect. \ref{data}.  A statistical analysis of the daily distributions is briefly recalled for completeness and coherence, repeating some information but also developing on  \cite{TWeffectPhA496.16.197}.
  in Sect. \ref{Discussion1};  correlations in daily submissions are examined in  Sect. \ref{Discussion2} in order to search for hidden structure as through "well known" but unusual  distributions (Sect. \ref{Weibull}). 
 An interference due to the editor role is discussed in Sect.\ref{editorrole}.   A possible seasonal influence is examined in Sect. \ref{season}.  A     Granger causality test  is also performed (Sect. \ref{Granger}). Since ARCH and generalized ARCH models might be unfamiliar to readers, in Sect.  \ref{ARCHstrong}, we explain the technical method leading to  an ARCH econometric-like model.
 In Sect. \ref{reasoning},  we interpret the model parameters and findings through the author's role at the manuscript submission time.  
The final section (Sect. \ref{conclusion}) re-emphasizes that the role of authors is the primary concern,  their behavior extracted from the data through an original modeling process.

  \begin{figure}
\includegraphics[height=10.8cm,width=10.8cm]{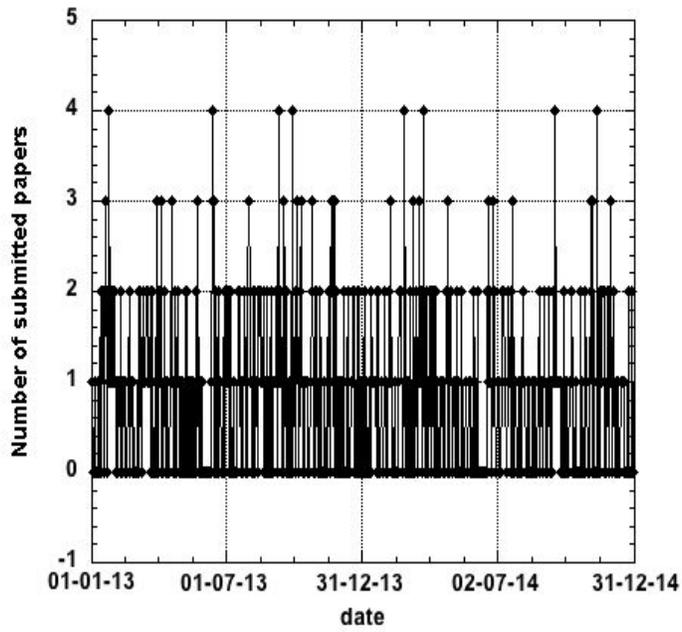} 
 \caption{
 Time series of the  number of papers submitted to JSCS according to the date of submission from Jan. 01,   2013 till Dec. 31,  2014.} \label{Plot13Ndayofweekjpeg} 
\end{figure}

  \begin{figure}
\includegraphics[height=8.8cm,width=10.8cm]{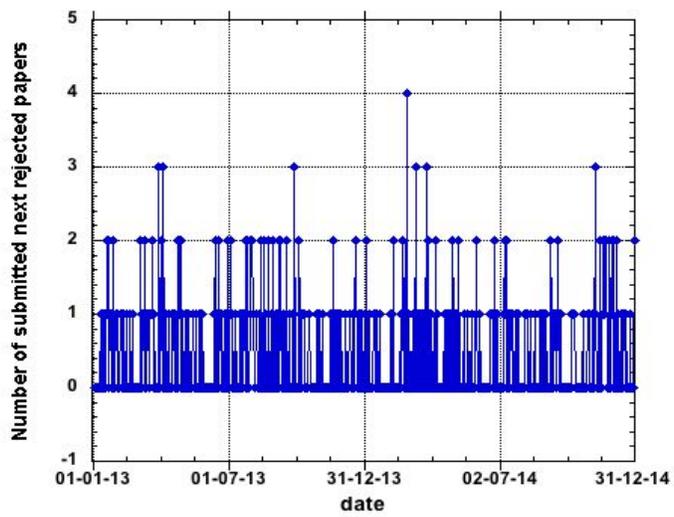} 
 \caption{
 Time series of the  number of papers submitted to JSCS according to the date of submission from Jan. 01,   2013 till Dec. 31,  2014, and subsequently rejected} \label{Plot12rtimeseriesjpeg} 
\end{figure}

\section{Data}\label{data}
Let us call  $N$   the number  of submitted paper on one given day,   the  later accepted papers ($N_a$) and those  later rejected ($N_r$). 
 (N.B.  In Table \ref{JSCSNNadaystat2},  $N\neq N_a+N_r \equiv N_s$, because  a few (4) papers were withdrawn by authors.)  The submission  time series daily records are  shown in Fig.  \ref{Plot13Ndayofweekjpeg}. 
 It can be mentioned that the periodogram gives a (huge) peak near  0.143 ($\simeq 2/7$).     A similar   graph could be displayed for $N_a$, but is not shown for saving space.    In fact, the latter could be also uploaded for any other  journal on internet. Recall that the number of submitted and the number of  rejected papers are not usually known (though see \cite{LearnPub24.11.325chemBornmann,LearnPub25.12.145psiSchreiber}, in cases concerned with seasonal effects). Therefore, and furthermore in view of the following discussion, it is original to display the  time series for the number of rejected papers when submitted on a given day. This is found in Fig. \ref{Plot12rtimeseriesjpeg}.
 
\begin{figure}
\includegraphics[height=10.8cm,width=12.8cm]{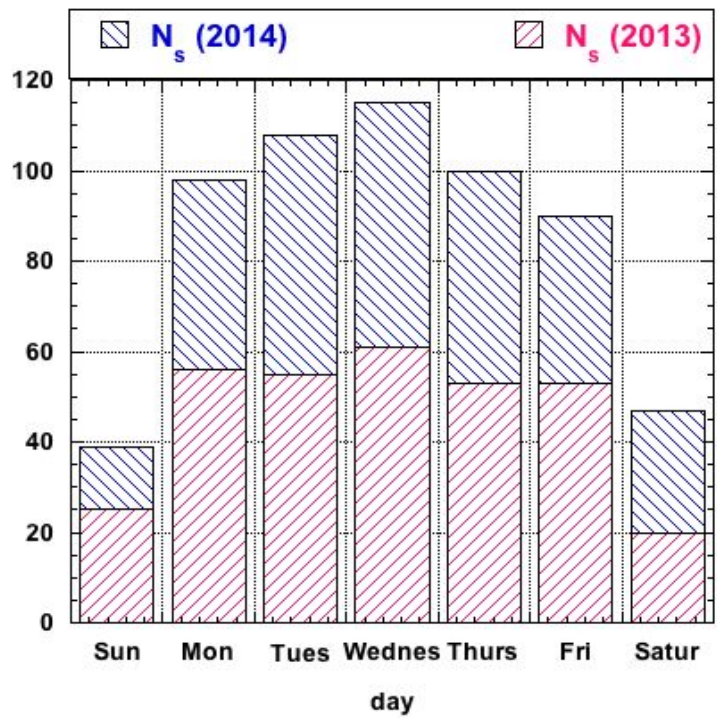} 
 \caption{$N_s$:  Number of papers submitted to JSCS according to the week day of submission in  2013 and 2014.} \label{Plot5S13S14histoday} 
\end{figure}

\begin{figure}
\includegraphics[height=8.8cm,width=12.8cm]{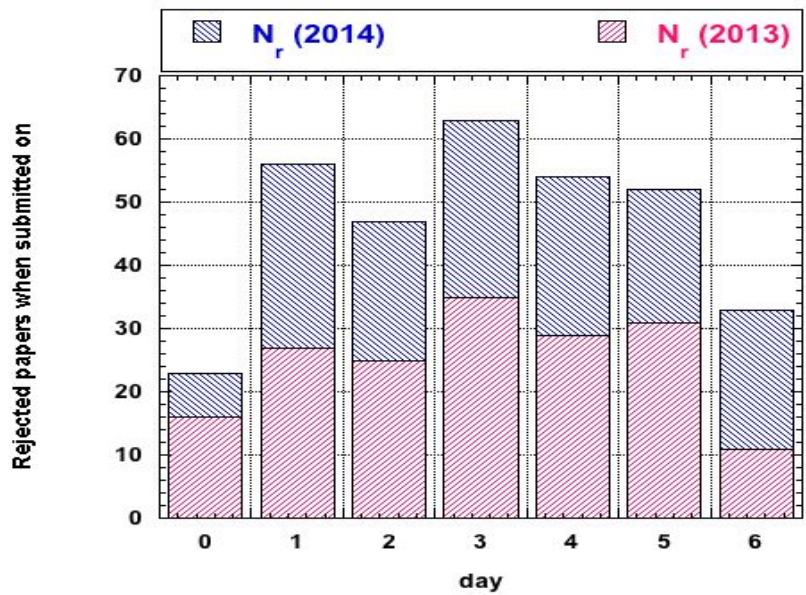} 
 \caption{ $N_r$: Number of rejected papers among those submitted to JSCS according to the week day of submission in  2013 and 2014;    Sunday (day 0), \dots,    Saturday (day 6).} \label{Plot7R13R14histoday} 
\end{figure}

\begin{figure}
\includegraphics[height=8.8cm,width=12.8cm]{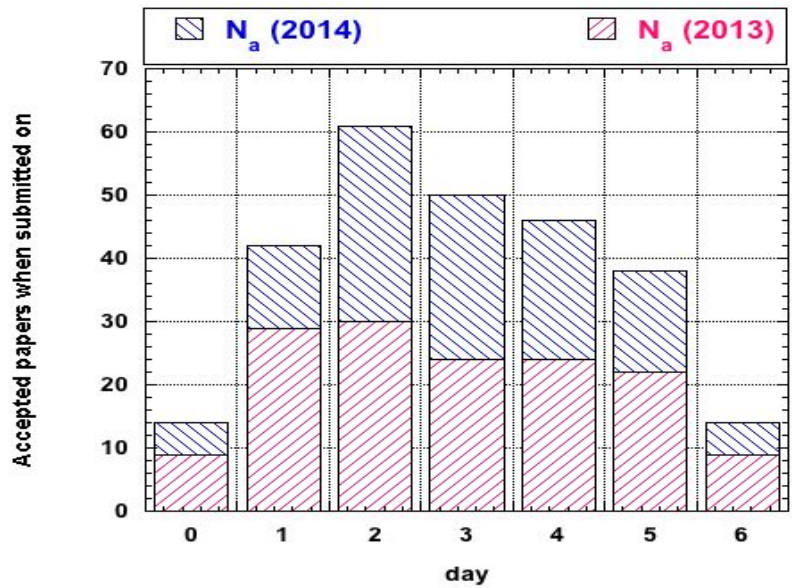} 
 \caption{ $N_a$: Number of accepted papers among those submitted to JSCS according to the week day of submission in  2013 and 2014;    Sunday (day 0), \dots,    Saturday (day 6).} \label{Plot6A13A14histoday} 
\end{figure}

 The histogram of the  number $N_s$ on a given week day   is shown in Fig. \ref{Plot5S13S14histoday}, emphasizing  both years of interest.  The corresponding histograms for the number of submitted next either rejected or accepted papers are given in Figs. \ref {Plot7R13R14histoday} - \ref{Plot6A13A14histoday} respectively.

\begin{figure}
\includegraphics[height=8.7cm,width=12.8cm]{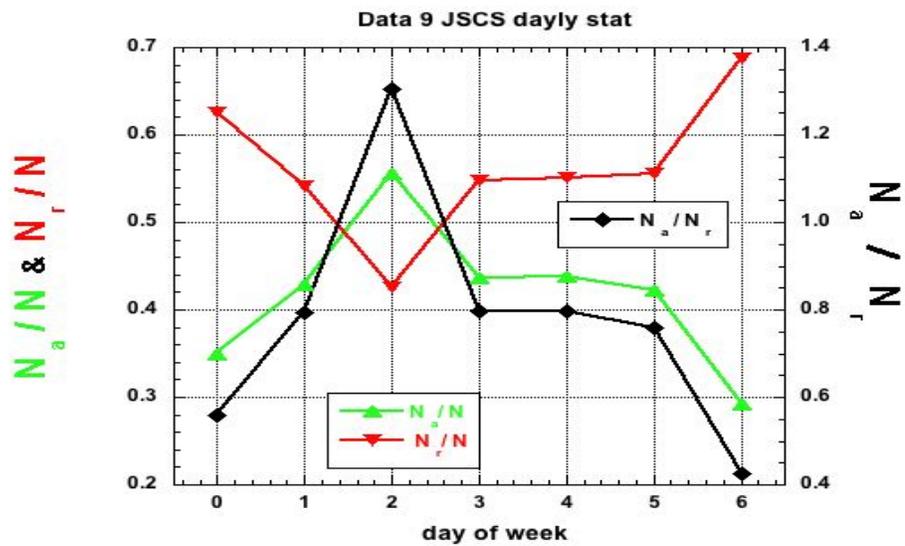}
 \caption{   Percentages of accepted  (triangle on basis) and rejected (triangle on tip)  papers, $N_a/N$ and $N_r/N$ respectively, according to the week day of submission  to JSCS   in  2013 and 2014;  diamonds: ratio of  accepted to rejected  papers  ($N_a/N_r$) according to the week day of submission to JSCS   in  2013 and 2014;    Sunday (day 0), \dots,    Saturday (day 6).} \label{Plot8NaNNrNNaNr} 
\end{figure}

    The relative number (expressed in percentages) of papers accepted or rejected after submission ($N_a/N_s$ and $N_r/N_s$)  on  a  specific day of the week is  given in Table \ref{JSCSNNadaystat2}. 
    The proportion of  (i) accepted to submitted papers ($N_a/N_s$), (ii) rejected to submitted papers ($N_r/N_s$), and (iii) accepted to rejected  papers  ($N_a/N_r$), according to  the day-of-the-week is shown in Fig.  \ref {Plot8NaNNrNNaNr}. The  ratio of  accepted to rejected  papers  ($N_a/N_r$) according to the week day of submission to JSCS   in  2013 and 2014 is also shown. The latter number can of course be greater than unity. This occurs on Tuesday.
  
   Thus, it appears that, in contrast to the more often occurring  submission day (Wednesday, weekday 3), - see Fig. \ref{Plot5S13S14histoday},  the papers are (relatively  with respect to the number of those submitted on the day) more often accepted when (or if) submitted on Tuesday (day  2). However, the largest number of rejected papers are those submitted on Wednesday (day  3), - see Fig. \ref{Plot7R13R14histoday}. When expressed in relative terms (in percentages of the submitted papers), - unexpectedly, the greatest proportion of manuscripts gets rejected if  they have been submitted on Sunday (day 0) or Saturday (day 6).

 \begin{table} \begin{center}
      \begin{tabular}{|c|c|c|c|c|c|c| c| c| }
      \hline      
      &	$N$	& 	$N_a$&	$N_r$& $N_a/N$& $N_r/N$&$N_a/N_r$ &$N_r/N_a$  \\
     \hline       \hline 
Min	&	0              	&	0              	&	0              	&	0              	&	0              	&	0              	&	0              	\\
Max	&	4              	&	3              	&	4              	&	1              	&	1              	&	3              	&	3              	\\
Sum	&	597 	&	265 	&	328 	&	177.08	&	212.92	&	102.67	&	97.833	\\
Mean	&	0.818	&	0.363 	&	0.449 	&	0.243	&	0.292	&	0.141	&	0.134 	\\
Std Dev	&	0.921 	&	0.594	&	0.678	&	0.391 	&	0.420	&	0.402	&	0.387 	\\
Std Err	&	0.0341	&	0.0220	&	0.0251	&	0.0145	&	0.01552 	&	0.0149	&	0.0143 	\\
Skewn	&	0.9788 	&	1.4865	&	1.4686	&	1.1808	&	0.9093	&	3.1567	&	3.3830	\\
Kurt	&	0.457 	&	1.522 	&	1.935	&	-0.3457	&	-0.969 	&	10.694	&	13.585	\\
\hline
$\chi^2$& 63.92& 50.884& 25.244& \multicolumn{4}{|c|}{[2013-2014]}  \\ \hline
$\chi^2$&35.127&21.524&17.414& \multicolumn{4}{|c|}{[2013] }  \\
$\chi^2$&32.365&36.00&14.545& \multicolumn{4}{|c|}{ [2014]}  \\
\hline
 \end{tabular}   \end{center}
\caption{ Statistical characteristics of the    submission  time series  (730 days: 2013-2014).   The characteristics of the percentage distributions are  also given.  The $\chi^2$ values are given for the whole time interval. They are  also given for each year of interest for comparison.}   \label{JSCSNNadaystat2}%
\end{table}

 
 \section{ Discussion of the Statistical Daily Distributions}\label{Discussion1}

Next, it appears to be of interest to verify whether these visual findings are statistically sound: a    $\chi^2$ value is naturally  in order, - see Sect. \ref{chi2}. Next,  in order to introduce an econometric-like, statistical, modeling, a test on the Weibull distribution seems  appropriate, - see Sect. \ref{Weibull}.

 \subsection{$\chi^2$ uniform distribution test}\label{chi2}
The $\chi^2$-test can only be made on the number of papers;  assuming a uniform daily distribution, one obtains values given on the last line of Table \ref {JSCSNNadaystat2}. They range from  $\simeq 14.5 $ till $\simeq  64.0$. Recall that the  $\chi^2$ value at 0.95\% confidence is  18.5476 for 6  degrees of freedom, for a uniform distribution 
 thereby indicating that the  submission distribution is far from uniform,  i.e.  there are significant differences about the day-of-the-week. There is a  markedly  significant propensity to submit on Wednesday (3rd  day-of-the-week).  The same conclusion can de drawn about   the going-to-be accepted papers.  The  later rejected papers are more uniformly distributed over the week, - when the yearly distributions are considered. However, when increasing the time range, the  $\chi^2$-test leads to a  $\simeq 25.0$ value, indicating a non-uniform distribution, in agreement with the visual perception  of the Fig. \ref{Plot7R13R14histoday}  data.

 Notice that  the distributions are rather skewed and the skewness  positive; the statistical characteristics are found in Table  \ref{JSCSNNadaystat2};  
 the kurtosis is also   negative. However an  asymmetry holds both  for the absolute values and for the percentages; for these the kurtosis is negative. Thus, the  characteristics of the various  distributions have next been examined in order to search whether they belong to a known case. 

\subsection{Other shape distribution tests} \label{Weibull}

Searching for   the  distribution shape, the Weibull probability  \cite{TFSC18.80.247SharifWeibull} was first considered; it was  originally imagined for survival    measures, - of humans or  electro-mechanical  goods. The Weibull failure model has also been applied in econophysics, for example, when  kinds of "first passage processes"  are relevant. {\it Mutatis mutandis}, it can be imagined that scientific papers are analog to light bulbs: the former survive if they are published, or die if they are rejected. They might be submitted and accepted elsewhere, of course.
 We have  performed a regression analysis for a "survival model" assuming that  "failure occurrences" have a Weibull distribution. By extension, we have also tested the Weibull distribution function on the submitted and later accepted papers.

The probability plot correlation coefficient 
 \cite{Filliben 1975}  is a graphical technique   for identifying the shape parameter for the distributional family that best describes the data set. Such a coefficient   has been looked for the 3 distributions ($N_s$, $N_a$, and $N_r$),  for the years 2013 and 2014 and for the whole time interval [2013--2014]. The plots are not "spectacular" and are not shown for saving space. 
 The optimal shape parameter is always found to be equal to 1, and exponentially decaying when increasing the number of shape parameters, suggesting that the  Weibull distribution can be adequate, - among the   simplest usually considered. Yet, the skewness and kurtosis seem "large" (see Table 1), demanding to further examine whether the "simplicity" of  the distributions  is so well demonstrated.

The maximum likelihood estimation 
 is another accurate and easy way to estimate distribution parameters \cite{MLE 8.4.1.2. Maximum likelihood estimation}. 
 The corresponding plots for  papers submitted and those subsequently  accepted or rejected on a given day present some structure. 
 The plots (not shown for saving space) do  not support normality, but the change in curvature (near 
  $\sim 0.13 \pm 0.03 $)
  and the curvature sign allow to  infer that the underlying distribution has a tendency toward a so called heavy tail,- whence suggesting a non dubious day effect, but hardly indicating a "classical distribution". We are aware that the number of days in a week  barely covers  one decade, - whence log-log plots for discovering the tail exponents appear to be  rather meaningless and are thus not shown. 
 
 Tukey tests \cite{Tukey 1949}  have also been performed: the resulting values  do not point to any of the most  usual distribution functions of random events; the distributions are found to be rather far from "exactly uniform".
 
 In conclusion of this subsection, it can be considered that the various day-of-the-week distributions do  not correspond to a usually well known distribution. 
 
  \section{
  Correlations in Daily Submissions}\label{Discussion2}
  
  Before attempting an econometric-like  regression model, it is necessary to appreciate the relevant variables. 
  One could consider that  the editor has a prominent role in rejecting papers, - at least, desk rejected ones. 
  However, there is no information on whether an editor does desk reject a paper on the submission day (or exactly on the corresponding 
  day of submission in another week). {\it A priori}, there should be no correlation between the day of submission and the day of desk rejection, since we hypothesize that the rejection follows a lack of quality of a paper because submitted on a given day.  Nevertheless  some  editor  role is briefly discussed in Sect. \ref{editorrole}.
   In the same spirit, one may wonder whether a seasonal effect can be observed. 
   Indeed most submitted papers are by authors belonging to some university or research laboratory. An academic time effect, 
   related to teaching load,  might be searched for. This question is tackled on Sect. \ref{season}.
  
 \subsection { Editor role possible influence} \label{editorrole}
 
 One might argue that the above distributions, in particular those about rejected papers, much depend on reviewers and editors behaviors,- the more so if there are desk rejected papers submitted on days during which the mood or duties of editors is not  "agreeable".  Notice that there were more than a dozen editors during the interval time relevant to this study. Moreover, editors do not necessarily examine submitted papers on the day of submission.  It occurred that a few editors were quite responsive, but one case of more than one month "delay in editor activity" could also been extracted from the data. 
 
The number of desk rejected papers $N_{dr}$ has been mentioned in \cite{TWeffectPhA496.16.197} 
to be  equal to 161.   Their day of submission distribution is quasi uniform  as depicted in Fig. 4 in \cite{TWeffectPhA496.16.197}  and  statistically proved through  a $\chi^2$  test  corresponding to the 95\% confidence of a null hypothesis (uniform distribution) for   6 degrees of freedom. Therefore, the week of day of submission for rejected papers does not seem to be editor dependent, whence allowing us not to consider such a variable for modeling the author behavior and the resulting outcome of his/her submission.

Nevertheless, for completeness,  it might be  interesting to find out  some editorial behavior,  through the frequency of desk rejection, in particular  the distribution of  the day of week when an editor desk reject a paper,  whenever the latter has been submitted. In Appendix, it is shown that such a distribution is markedly different from that of the author submission behavior and from that of the  distribution of days of submission for rejected papers.  This allows us to reject an editorial effect about the submission  of papers,  and about the outcome (acceptance or rejection) of submitted papers, whence such a variable will not be considered in the following model.
 
 \subsection{ Possible seasonal influence} \label{season}
 
\begin{figure}
\includegraphics[height=8.8cm,width=12.8cm]{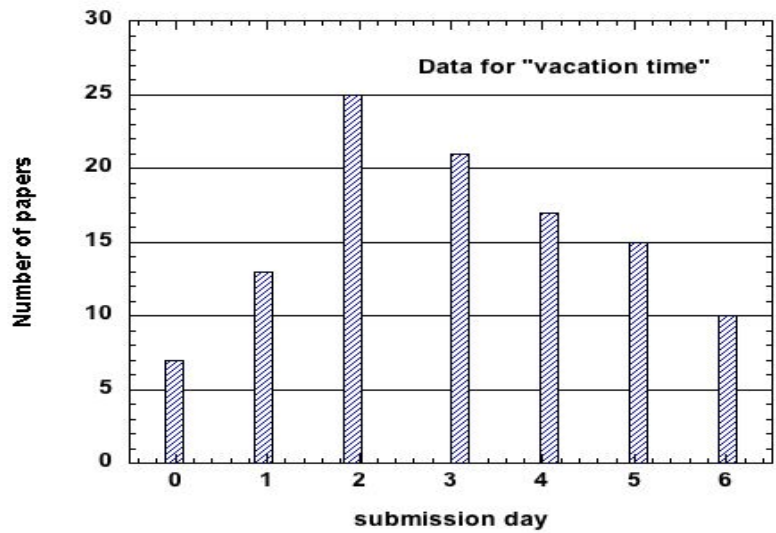} 
 \caption{  Number of submitted  papers  on a given day of the week   during so called vacation time (July and August) in  2013 and 2014;    Sunday (day 0), \dots,    Saturday (day 6).} \label{Plot7submdayvac} 
\end{figure}

 One might argue that authors have a different schedule in the teaching  months than during vacation time (say, somewhat arbitrarily between July 01 and August 31).  Thus, it  may be usefully tested whether the anomalous distribution of submitted papers and of those accepted (or rejected) depends on the year timing. If such a "teaching duty during academic  time" is relevant, it might be expected that the distribution is more uniform during vacation times. We selected the data for those time intervals.  The cumulative number of submitted  papers  for both years  on a given day of the week   during such so called vacation time (July and August)  is shown on Fig. \ref{Plot7submdayvac}.  A peak occurs on Tuesday, and a small number of papers is submitted during weekend days. The distribution is visually far from uniform. Thus, it seems apparent that the submission pattern does not depend much on the season, i.e. on academic duties.
 
\begin{figure}
\includegraphics[height=8.8cm,width=10.8cm]{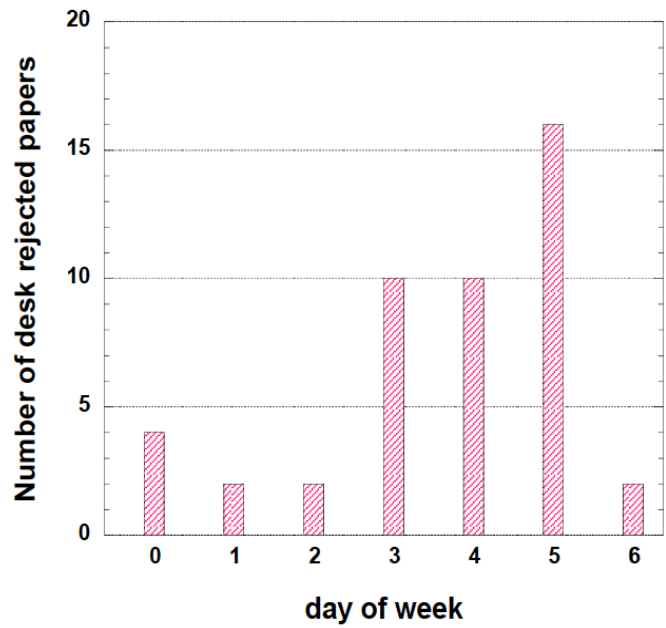} 
 \caption{  Number of desk rejected papers ($N_{dr}$)  by  editors  on a given day of the week among those submitted to JSCS  whatever the week day of submission during the months of july and august in  2013 and 2014;    Sunday (day 0), \dots,    Saturday (day 6).} \label{Plot4rejdayvac} 
\end{figure}

  The  overall acceptance and rejection fates during vacation time are not examined because they depend on reviewers more than on authors; indeed the paper fate can occur  at a later  time than July or August. Nevertheless, the desk rejection can be examined to emphasize the editor role, if any, during such a holiday time.  This is shown on  Fig. \ref{Plot4rejdayvac}.  As for the academic year, the number of submitted but desk rejected papers increases  toward the end of the week.  This might imply some effect due to, or correlation with, the mood of editors,  during vacation time or before a weekend.
  
   \subsection{Granger Causality Test  } \label{Granger}

  In view of the above, and with the aim of searching for a model of behavior, one is geared toward further considering
   the absence of correlations between daily submissions by the different agents. A Granger test of causality seems in order. 
   The Granger causality test is a statistical hypothesis test for determining whether one time series is useful in forecasting another. 
   This is   a classical test, for example, systematically  applied to the returns on    stock indices  \cite{DEM14.14.93}.
    Thus, it  can be considered  as an {\it ad hoc}  tool for outlining a restricted choice of explanatory variables in the subsequent modeling.  
The Granger test of  causality is based on the vector auto regression (VAR)  type regressions, i.e. 
 regression of $Y$, i.e. a vector $\equiv y_t$, on its lagged values and the same lags of the $X$, a vector $\equiv x_t$,  variable:
 \begin{equation}\label{DEMeq3} 
  y_t= a_{1,1} y_{t-1} + .... +a_{1,k} y_{t-k}+  b_{1,1} x_{t-1} +... + b_{1,k} x_{t-k} + \epsilon_t
 \end{equation}
The null hypothesis $H_0$  corresponds to  $b_{1,j}=0$ for all $j$.  If so, it means that the $X$ does not "Granger-cause"
the $Y$ variable.  
 
Derived through  "ordinary" least square method (OLS) estimates with robust standard errors,  typical  results of the Granger causality tests for   causality correlations between $N_s$, $N_a$, and $N_r$   are given in Table \ref{Granger7},  
 based on F-test statistics values; $p$-values are given. 
 
It can be deduced  that the null hypothesis on  lack of causality is rejected for all pairs of  events  with  the exception of  $N_a$ = f($N_r$).  The time series  are independent  of each other (as  it should  be somewhat  expected, it seems). Other time lags have been  tested, but have not shown to be  carrying  any surprising information (bis).

 \begin{table} \begin{center}
      \begin{tabular}{|c|c|c|c| }
      \hline      
      Model	&	 	F	&	p-value	\\	  \hline   
        $N_a$ = f($N_s$)	& 1.6902	&	0.1082	\\				 			
              $N_s$ = f($N_a$)	& 0.3999	&	0.9026	\\ \hline   		 					
              $N_r$= f($N_s$)	& 0.6622	&	0.7042	\\		 			 		
              $N_s$= f($N_r$)	& 0.4493	&	0.8709	\\		\hline   			
              $N_r$ = f($N_a$)	& 0.8249	&	0.5667	\\		 	 			
 $N_a$ = f($N_r$)	& 1.8717	&	0.0714¬*¬*¬*	\\			\hline
 \end{tabular}   \end{center}
\caption{  Granger test typical results \cite{Wessa 2013,Seth 2007}. The maximum number lags (of the endogenous variable) that is here used in the test equation was specified equal to -7. Reduced model = 715 days; complete model = 708 days.  The *** conventionally indicate a 10\% level significance only; all other p-values  are 5\% significant.   
}   \label{Granger7}%
\end{table}

 \section{Daily Submission Econometric-like Modeling}\label{Discussion}
 
  Therefore, this paper adopts the forms of  
so called    strong modeling  of the day-of-the-week effects of returns ($r_t$) on a stock market, through the ARCH (Autoregressive Conditional Heteroskedasticity) methodology, but differs in the input time series.  For the unfamiliar reader, let it be recalled that in  econometrics, a "weak model"  has one dummy variable only, so that the time series regression only compares the significance of different coefficients   \cite{bollerslev1994arch}. For example, $ r_t = \alpha_0 + \alpha _1     M_t + e_t$, where  $M_t$ represents the value of the time series on a given Monday; $\alpha_0$ and $\alpha_1$ are  to be determined.
  
  In  an econometrics  "strong model", one  compares the difference of return rates on one day with those of the  other (four) trading days.  For example
  \begin{equation} \label{strongeqmonday}
r_t ^{Mo}=  \gamma_0 +  \gamma_2\: Tu  +   \gamma_3  \;We  +  \gamma_4\; Th  +  \gamma_5\; Fr  +  \Gamma_1,
\end{equation}
where the coefficients  are the unknowns to be determined.  Thus, the  day-of-the-week effect in  a "strong test" in econometrics refers to the yield rate in a trading day, searching whether  this rate is  significantly higher or lower than on any other   trading day. In this case, five regression equations, like Eq.(\ref{strongeqmonday}) are written to   determine the relative size of the 
yield rate in any  of the (five) trading days      \cite{brennan1996market}. 

A standard OLS method   is used for the regression analysis over the single or multiple virtual variables.  
In bibliometrics, it seems inconvenient to  us  "return" as the appropriate word  for measuring  the fate of scientific papers. The number of these can hardly be called a "price", as on a stock market, - although authors when submitting a paper, in some sense, try  "to sell" it to the editor, next to the reviewers, and finally to the community.  Thus the analogy with econometrics stops here at the methodology. In brief, the "short model" will consider  one single time series to be regressed through the \underline{seven}  day time series. One will obtain seven coefficients which are to be discussed for observing some significance or not. In contrast, the "strong model" will consider  \underline{seven}  reduced time series, - each one being the original time series measured with respect to  the average number (in the appropriate time interval) of submissions in one of each day of the week.

  \subsection{  ARCH-like   model} \label{ARCHstrong}

  \begin{table} \begin{center}
      \begin{tabular}{|c|c|c|c|c|c|c|c|c| }
      \hline      
       $\beta_k$	 &   \multicolumn{7}{|c|}{[day  time series]}   \\   
     \hline      
       $k=$ & $Sun$ & 	$Mon$&	$Tue$& $Wed $& $Thu$&$Fri$& $Sat$ \\
     \hline    
       &   \multicolumn{7}{|c|}{[2013-2014]} \\  \hline 
0	 &  0.0000 &   -0.5664 &   -0.6525 &   -0.7381 &   -0.5664 &   -0.4896 &   -0.0768\\
 1&  0.5664  &  0.0000 &   -0.0861 &   -0.1717  &  0.0000  &  0.0768  &  0.4896\\
2&  0.6519  &  0.0860 &   -0.0000 &   -0.0855  &  0.0860  &  0.1628  &  0.5752\\
 3  &  0.7374  &  0.1715  &  0.0855 &   -0.0000  &  0.1715  &  0.2483  &  0.6607\\
 4 &  0.5664  &  0.0000 &   -0.0861 &   -0.1717  &  0.0000  &  0.0768  &  0.4896\\
 5 &   0.4896 &   -0.0768 &   -0.1629 &   -0.2485 &   -0.0768 &   -0.0000  &  0.4128\\
6  &    0.0768 &   -0.4896 &   -0.5757 &   -0.6613 &   -0.4896 &   -0.4128  &  0.0000\\ \hline					
								  &   \multicolumn{7}{|c|}{[2013]}  \\ \hline 																																					
0	 &  -0.0000 &  -0.5952 &  -0.5561 &  -0.6912 &  -0.5376 &  -0.5376 &   0.0960\\
 1&   0.5952 &   0.0000 &   0.0391 &  -0.0960 &   0.0576 &   0.0576 &   0.6912\\
 2&   0.5579 &  -0.0392 &   0.0000 &  -0.1356 &   0.0185 &   0.0185 &   0.6542\\
 3&   0.6912 &   0.0960 &   0.1351 &  -0.0000 &   0.1536 &   0.1536 &   0.7872\\
 4&   0.5376 &  -0.0576 &  -0.0185 &  -0.1536 &   0.0000 &   0.0000 &   0.6336\\
 5&   0.5376 &  -0.0576 &  -0.0185 &  -0.1536 &   0.0000 &   0.0000 &   0.6336\\
 6&  -0.0960 &  -0.6912 &  -0.6520 &  -0.7872 &  -0.6336 &  -0.6336 &  -0.0000\\ \hline					
								  &   \multicolumn{7}{|c|}{[2014]} \\  \hline 					
0& 0.0000  & -0.5376  & -0.7488  & -0.7861  & -0.5952  & -0.4416  & -0.2496\\
1&    0.5376&    0.0000  & -0.2112  & -0.2485  & -0.0576&    0.0960&    0.2880\\
2 &    0.7488&    0.2112&    0.0000  & -0.0373&    0.1536&    0.3072&    0.4992\\
3 &    0.7887&    0.2493&    0.0375&    0.0000&    0.1915&    0.3457&    0.5383\\
4 &    0.5952&    0.0576  & -0.1536  & -0.1909&    0.0000&    0.1536&    0.3456\\
5 &    0.4416  & -0.0960  & -0.3072  & -0.3445  & -0.1536&    0.0000&    0.1920\\
6&    0.2496  & -0.2880  & -0.4992  & -0.5365  & -0.3456  & -0.1920&    0.0000\\
\hline
 \end{tabular}   \end{center}
\caption{ $\beta_k$ values  of the    submitted paper  time series    (730 days in 2013-2014; 365 days in  either 2013 or 2014). 
}   \label{betaNs}%
\end{table}

   \begin{table} \begin{center}
      \begin{tabular}{|c|c|c|c|c|c|c|c|c| }
      \hline      
       $\beta_k$	 &   \multicolumn{7}{|c|}{[day  time series]}   \\   
     \hline      
       $k=$ & $Sun$ & 	$Mon$&	$Tue$& $Wed $& $Thu$&$Fri$& $Sat$ \\
     \hline    
       &   \multicolumn{7}{|c|}{[2013-2014]} \\  \hline 
0	       &  -0.0000 &  -0.2688 &  -0.4456 &  -0.3600 &  -0.2880 &  -0.2304 &  -0.0000\\
1  &  0.2688 &  -0.0000 &  -0.1768 &  -0.0912 &  -0.0192  &  0.0384  &  0.2688\\
 2 &  0.4452  &  0.1767 &  -0.0000  &  0.0855  &  0.1575  &  0.2150  &  0.4452\\
 3 &  0.3597  &  0.0912 &  -0.0855  &  0.0000  &  0.0720  &  0.1295  &  0.3597\\
 4 &  0.2880  &  0.0192 &  -0.1576 &  -0.0720 &  -0.0000  &  0.0576  &  0.2880\\
 5 &  0.2304 &  -0.0384 &  -0.2152 &  -0.1296 &  -0.0576  &  0.0000  &  0.2304\\
 6&  -0.0000 &  -0.2688 &  -0.4456 &  -0.3600 &  -0.2880 &  -0.2304 &  -0.0000\\ \hline					
		  &   \multicolumn{7}{|c|}{[2013]}  \\ \hline 																																					
0	  &  -0.0000 &  -0.3840 &  -0.3923 &  -0.2880 &  -0.2880 &  -0.2496 &  -0.0000\\
 1&   0.3840 &   0.0000 &  -0.0083 &   0.0960 &   0.0960 &   0.1344 &   0.3840\\
 2&   0.3936 &   0.0084 &   0.0000 &   0.1047 &   0.1047 &   0.1432 &   0.3936\\
 3&   0.2880 &  -0.0960 &  -0.1043 &   0.0000 &   0.0000 &   0.0384 &   0.2880 \\
 4&   0.2880 &  -0.0960 &  -0.1043 &   0.0000 &   0.0000 &   0.0384 &   0.2880\\
 5&   0.2496 &  -0.1344 &  -0.1427 &  -0.0384 &  -0.0384 &  -0.0000 &   0.2496\\
 6&  -0.0000 &  -0.3840 &  -0.3923 &  -0.2880 &  -0.2880 &  -0.2496 &  -0.0000\\ \hline					
								  &   \multicolumn{7}{|c|}{[2014]} \\  \hline 																														
0	  &   0.0000 &   -0.1536 &   -0.4992 &   -0.4315 &   -0.2880 &   -0.2112 &   0.0000\\
 1&   0.1536 &   0.0000 &   -0.3456 &   -0.2779 &   -0.1344 &   -0.0576 &   0.1536\\
 2&   0.4992 &   0.3456 &   -0.0000 &   0.0677 &   0.2112 &   0.2880 &   0.4992\\
 3&   0.4329 &   0.2788 &   -0.0680 &   0.0000 &   0.1439 &   0.2210 &   0.4329\\
 4&   0.2880 &   0.1344 &   -0.2112 &   -0.1435 &   -0.0000 &   0.0768 &   0.2880\\
 5&   0.2112 &   0.0576 &   -0.2880 &   -0.2203 &   -0.0768 &   0.0000 &   0.2112\\
 6&   0.0000 &   -0.1536 &   -0.4992 &   -0.4315 &   -0.2880 &   -0.2112 &   0.0000  \\  \hline 	
 \end{tabular}   \end{center}
\caption{ $\beta_k$ values  of the accepted paper  time series    (730 days in 2013-2014; 365 days in  either 2013 or 2014). 
}   \label{betaNa}%
\end{table}

   \begin{table} \begin{center}
      \begin{tabular}{|c|c|c|c|c|c|c|c|c| }
      \hline      
       $\beta_k$	 &   \multicolumn{7}{|c|}{[day  time series]}   \\   
     \hline      
       $k=$ & $Sun$ & 	$Mon$&	$Tue$& $Wed $& $Thu$&$Fri$& $Sat$ \\
     \hline    
       &   \multicolumn{7}{|c|}{[2013-2014]} \\  \hline 
0	        &  0.0000 &   -0.3168 &   -0.2261 &   -0.3782 &   -0.2976 &   -0.2784 &   -0.0960\\
1  &  0.3168 &   -0.0000  &  0.0907 &   -0.0614  &  0.0192  &  0.0384  &  0.2208\\
 2 &  0.2259 &   -0.0906 &   -0.0000 &   -0.1520 &   -0.0714 &   -0.0523  &  0.1300\\
 3 &  0.3779  &  0.0614  &  0.1520  &  0.0000  &  0.0806  &  0.0997  &  0.2820\\
 4 &  0.2976 &   -0.0192  &  0.0715 &   -0.0806  &  0.0000  &  0.0192  &  0.2016\\
 5 &  0.2784 &   -0.0384  &  0.0523 &   -0.0998 &   -0.0192  &  0.0000  &  0.1824\\
 6 &  0.0960 &   -0.2208 &   -0.1301 &   -0.2822 &   -0.2016 &   -0.1824 &   -0.0000\\ \hline					
		  &   \multicolumn{7}{|c|}{[2013]}  \\ \hline 																																					
0   &   0.0000 &  -0.2112 &  -0.1637 &  -0.3648 &  -0.2496 &  -0.2880 &   0.0960\\
 1&   0.2112 &   0.0000 &   0.0475 &  -0.1536 &  -0.0384 &  -0.0768 &   0.3072\\
 2&   0.1643 &  -0.0476 &  -0.0000 &  -0.2017 &  -0.0861 &  -0.1247 &   0.2606\\
 3&   0.3648 &   0.1536 &   0.2011 &  -0.0000 &   0.1152 &   0.0768 &   0.4608\\
 4&   0.2496 &   0.0384 &   0.0859 &  -0.1152 &   0.0000 &  -0.0384 &   0.3456\\
 5&   0.2880 &   0.0768 &   0.1243 &  -0.0768 &   0.0384 &  -0.0000 &   0.3840\\
 6&  -0.0960 &  -0.3072 &  -0.2597 &  -0.4608 &  -0.3456 &  -0.3840 &   0.0000\\ \hline				
								  &   \multicolumn{7}{|c|}{[2014]} \\  \hline 																														
0	&   -0.0000 &   -0.4224 &   -0.2880 &   -0.3931 &   -0.3456 &   -0.2688 &   -0.2880\\
1 &   0.4224 &   0.0000 &   0.1344 &   0.0293 &   0.0768 &   0.1536 &   0.1344\\
2 &   0.2880 &   -0.1344 &   -0.0000 &   -0.1051 &   -0.0576 &   0.0192 &   -0.0000\\
3 &   0.3944 &   -0.0294 &   0.1054 &   0.0000 &   0.0476 &   0.1247 &   0.1054\\
4 &   0.3456 &   -0.0768 &   0.0576 &   -0.0475 &   -0.0000 &   0.0768 &   0.0576\\
5 &   0.2688 &   -0.1536 &   -0.0192 &   -0.1243 &   -0.0768 &   0.0000 &   -0.0192\\
6 &   0.2880 &   -0.1344 &   -0.0000 &   -0.1051 &   -0.0576 &   0.0192 &   -0.0000\	\\ \hline
 \end{tabular}   \end{center}
\caption{ $\beta_k$ values  of the   rejected paper time series    (730 days in 2013-2014; 365 days in  either 2013 or 2014). 
}   \label{betaNr}%
\end{table}
  
 The "modeling" starts from 7 equations,  each  left hand side being a  time series describing the number of events (which has occurred on a given day).
  Under a matrix equation  form, the system of equations reads
 \begin{equation}\label{eqDY}
 \Delta Y \equiv Y  -<Y > =  X .\beta  + \epsilon
\end{equation}
where $Y $ is a (vector)  time series  and  $ <Y > $  the  corresponding average value of an event  (in the appropriate time interval), both  for some  (the same)  given week day; $X$ is a  ($t_M$ x $k$) rectangular matrix defining  the day when   an  event occurred, while the $\epsilon $ vector is supposed to be a white noise;  the (7) components of the  $\beta$ vector have to be determined.
In scalar notations,  one has
\begin{equation}\label{eqDYtk}
 Y_{t,k}\equiv Y_t -<Y_t>_k =  X_ {t,k} \beta_k + \epsilon_t 
\end{equation}
where  $t$ denotes the  successive days in the time series: in our case  $t \in  [1,t_M]$ where  $t_M$= 730 days in 2013-2014, but 365 days in  either 2013 or 2014. The index $k$  indicates the day of the week:
  $k\in[0,6]$.  In other words,
  \begin{equation}\label{eqMo}
   \Delta Y_t ^{Mon}\simeq  \beta_0 \;Sun  +   \beta_1 \;Mon  +   \beta_2 \;Tue  +   \beta_3 \;Wed  +  \beta_4 \;Thu  +  \beta_5 \;Fri  +  \beta_6 \;Sat, 
  \end{equation} 
  where, for example,  $Sun$ is a vector which has components equal to 1 each Sunday of the interval  and 0 otherwise. The best estimation  for the regression parameters $\beta_k$  holds through  the  Ordinary Least Squares  method for the sum of the "error" squares:
     \begin{equation} \label{S2}
\mathbf{S} \equiv \Sigma e_t^2
\end{equation} 
  minimizing $ \mathbf{S}$ with respect to $\beta$, i.e. let the first derivative equal to 0, one gets: 
    \begin{equation}\label{eqbeta}
\beta =  (\tilde{X}\;X)^{-1}  \;\tilde{X} \;  \Delta Y 
\end{equation}where $ \;\tilde{X}$ is the transposed of $X$.

The  $\beta_k$ coefficients  of the    submission  time series are  given for each year of interest  and (for comparison) for the whole time interval in Table \ref{betaNs}.   The corresponding values of  the papers submitted on a given day and either (later) accepted or (later)  rejected 	are given in Table \ref{betaNa} and Table \ref{betaNr}, respectively.   The data is given with 4 decimals:  the last one points to the (expected) precision in  the regression parameter estimation. 
 For the reader's ease, let it be made clear that, for example,  the $\beta_k$ value $ - 0.5664$,  in Table \ref{betaNs},  corresponds to $\beta_0$ in Eq.(\ref{eqMo}). The matrix of $\beta$'s as given in the Tables is of course antisymmetric, whence the "diagonal" has 0.  Sometimes  $\beta_k=0$ outside the "diagonal".  This is a consequence of the fact that the mean number of events occurs to be the same on two different days. The $\beta_k$  sign indicates whether  the contribution is "positive" or "negative" with respect to the mean of the day.  It can be noticed that the sign can change from a time interval to another, but this (of course) only occurs for small values of $\beta_k$.  
 
 The differences between the various $\beta$'s indicate the relevance of a week day with respect to another  one.    Recall that if  $ \beta_k$  is significantly different from 0, it can be considered that the   event ("yield rate" in econometrics)  is  significant as  compared with the other days, namely there is an "effect" on a given day, - due to activity (or lack of activity) in other days.  Visually, the relevance of the days in the middle of the week is noticeable and well reproduced for the submission and (later) acceptance of papers. In  contrast, the most likely rejection of papers submitted on the weekend is less systematic. This mathematically illustrates the skewed distribution shown and seen in Fig. \ref{Plot7R13R14histoday}.


\subsection{Reasoning}\label{reasoning}
Recall that we propose the alternative to  the null hypothesis  about the significance of days of submissions of papers submitted to a scientific journal. We propose that  the quality of submitted papers is somewhat reflected through the day of submission.
\begin{itemize} \item
Indeed, it can be conjectured that researchers who submit on Saturday or Sunday do so   because they want top get  rid of papers;  some  pressurizing coauthor or themselves expect the manuscript to have been sent by Friday. Maybe, these scientists are less eager to read very carefully once again their manuscript or  are less inspired to redraw   conclusions or  to demonstrate  significant relations in the results or to report to their boss that  "things" have not yet been finalized. 

\item 
 A comment on the  ARCH regression model is in order here in  view of  distinguishing it from "agent based models".  By analogy with econometric forecasters who have found some possibility to predict future variations, it can be  immediately suggested that the ARCH model provides one way of forecasting submission variance   change over time, outside (or beyond)  the usual  Markov memory free process. 
  Thus, this type of statistical model  has a variety of characteristics which should make it attractive for bibliometrics and scientometrics  applications, but also allow technological means of artificial intelligence for helping editors \cite{MFFNAeditwork}. 
  
  Nevertheless,   the ARCH model can be criticized because it assumes that the variance does not change with time. However, this does not seem   to be a strong assumption here. Indeed one can verify that the  [2013-2014] variance = 0.8488, while the [2013] and [2014] are  respectively equal to 0.8933  and  0.7976, - for submitted manuscripts on a given day.   In order to take into account a  time dependent variance, a GARCH model  \cite{GARCHmodelEngle} would be in order, but this is outside the scope of the present paper.
 \end{itemize}

\section{Conclusion}\label{conclusion}

First, in summary, there is quite a number of studies on the "day-of-the-week" effect on financial markets. To the best of our knowledge, this is the first time that one quantifies the submission of scientific papers, thus the behavior of scientific agents in such a process, i.e. considering some author's brain work and scientific activity content, depending on the day-of-the-week.
  It has   been shown that the analysis should take into account the relative size of daily submissions within a week. This normalization is relevant in order to observe whether the acceptance and rejection rates will differ depending on the day of submission.  In view of the high relative independence of scientific agents submitting papers, a statistical analysis of the submission time series is the most appropriate one. This has been attempted through an econometric formulation like the ARCH model.

Interestingly, since the available data gives some information on the fate of each submitted paper, acceptance or rejection, these two time  series have been statistically and "econometrically" analyzed. It is concluded that in the examined case, there is a significant middle of the week effect, not only for the submission, but also for the acceptance rate. It is surmised that the quality of submitted papers varies with the day of submission. Some explanation of the finding is given in terms of stress and collective pressure. It is true that such a hypothesis would be better confirmed if a study of the number of authors and their expertise level was added to the analysis. This would be an easy addition of a couple  terms ("variables") in the ARCH modeling, but the matter falls far outside the present aims.

Within the present framework and methodology, another aspect could be interestingly examined, i.e. the  "relative time series": $N_aN_s$, $N_r/N_s$, and $N_a/N_r$.  Indeed the conclusion based on absolute numbers and relative numbers could be debated, - the more so depending on the denominator in an expression "relative to what". Thus, an ARCH modeling of percentages might allow some further discrimination in the \underline{relative} importance of the correlations in  the day-of-the week effect.  

In conclusion, let  us to offer a few suggestions for further research lines, first based on possible "ambiguities" in our findings. It would be interesting to see: (i) whether there is a general trend in authors' behavior when choosing the submission day of the week and (ii) whether the final decision on the paper exhibits a similar relation to the day of submission in other types of  scientific papers. According to our results, it seems that weekend days (Saturday and Sunday) are not the best time for  finalizing and submitting manuscripts. We have been intrigued to see how many papers are desktop rejected.  The role of editor does not seem to be stochastic; it confirms the hypothesis that the apparent quality of a paper depends is correlated to the day of submission.

Of course, we admit that the fate of  a manuscript  depends on many  peer review process participants, beside the authors,   \cite{Hargens}, namely editors and reviewers  \cite{SCIMpeerZipf}. In order to reveal (possible) "day-of-the-week" effect in the entire process of scientific publication, it would be of interest to investigate when reviewers are informed that they should review a paper, when they accept (or refuse) to review the submission, when they comment on the paper,  etc., 
but such data for the JSCS are alas not available. However, these data rather pertain to reviewers and editors behaviors, - not to authors, as focussed upon here.
 
  \vskip0.5cm
 {\bf Acknowledgements}   \vskip0.5cm
  This paper has been part of scientific activities in COST Action  TD1306 New Frontiers of Peer Review (PEERE).

MA thanks  Sonia Bentes,  Roy Cerqueti, Claudiu Herteliu,  
Bogdan V.  Ileanu, and Claudio Lupi  for comments and advices on  the ARCH methodology.

MM,  PF, and AF were supported by internal funds of the Faculty of Physics at Warsaw University of Technology.
 \vskip0.5cm
 \clearpage
 {\bf Appendix} \vskip0.5cm {\bf Note on the day of  desk rejection} \vskip0.5cm

\begin{figure}
\includegraphics[height=7.8cm,width=12.8cm]{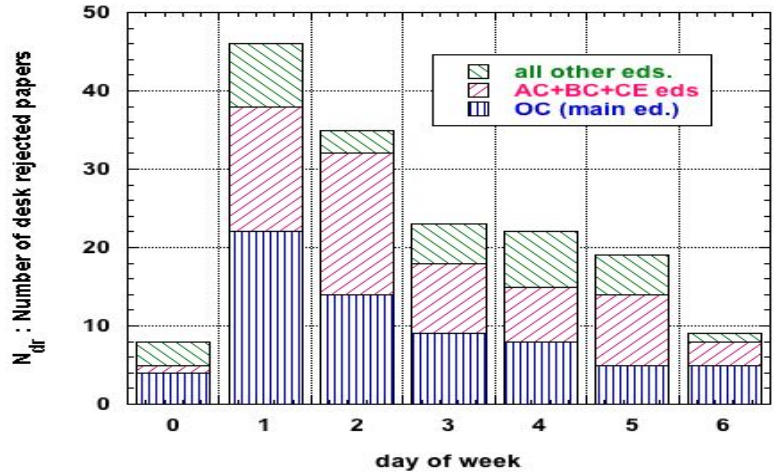}
 \caption{  Number of desk rejected papers ($N_{dr}$)  by   editors, selected by duty load,   on a given day of the week for  papers submitted to JSCS  whatever the week day of submission in  2013 and 2014;    Sunday (day 0), \dots,    Saturday (day 6).} \label{Plot2dayofweekrejectiony}  
\end{figure}

 In this Appendix, we examine the   editorial behavior with respect to the rejection of papers. Although it is unlikely that an editor 
 waits an exactly multiple of 7 for examining a submitting paper and desk rejecting it, one might  in so doing  observe when editors are more active during 
 specific days of the week or not, and whether their rejection rate  has a specific pattern related to the aper submission. In order to do so, we have examined on what day of 
 the week a paper is desk rejected. Of course, this is strictly an editor affair, unknown to the submitting author. For such a query,  and display ease,
  we have made 3 groups of editors among the dozen or so in  charge of subfields: the "main editor"  carrying about  40\% of the work load, i.e., 
  the  so called OC (organic chemistry subfield), the next  editors (analytical chemistry, biochemistry, and chemical engineering, - 
 in short   AC, BC, and CE subfields), and the other  10 or so editors. The number of desk rejected papers on a given day of the week, whenever the paper was submitted is shown in    Fig. \ref{Plot2dayofweekrejectiony}. It can be observed that the distribution is an exponentially decaying function ($R^2\simeq 0.97$)  of the day of the week, with a "relaxation time" $\simeq  3.460$.  
 
 Apparently editors seem very active at the beginning of the week for rejecting papers, we insist, whatever their submission day. It seems that one can be easily convinced that  editors (we do not have data for the reviewer recommendation day) are equally moody and  fair (or unfair) during the whole week irrespective of the day;  see related discussions in \cite{SpringerPlus5.16.903Wang} and  \cite{PONE6.13.e85382lortie}. Thus,  such an {\it a posteriori} effect, unknown to authors has not to be included in the relevant variables of the model.
  
This number of desk rejected papers on a given  day (Fig. \ref{Plot2dayofweekrejectiony}) can be usefully compared to data in  
Fig. \ref{Plot5S13S14histoday} and  Fig. \ref{Plot7R13R14histoday} pertaining to the submission day and rejection fate depending on the submission day.
From these, the related percentages can be easily deduced for comparison to data in Fig. \ref{Plot8NaNNrNNaNr}. This is left for the reader perusal.

\begin{figure}
\includegraphics[height=7.8cm,width=12.8cm]{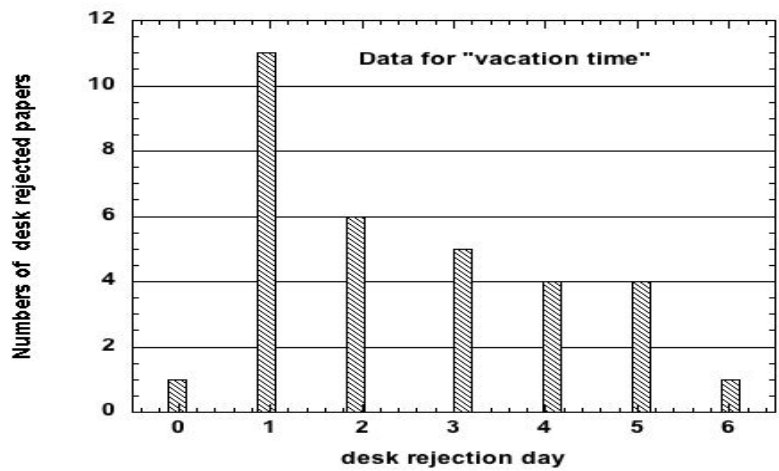} 
 \caption{  Number of desk rejected papers ($N_{dr}$)  by (all) editors  \underline{during so called vacation time} (July and August) in  2013 and 201  on a given day of the week for papers submitted to JSCS  whatever the week day of submission4;    Sunday (day 0), \dots,    Saturday (day 6).} \label{Plot9rejdayvac} 
\end{figure}

 For completeness,   in so doing adding to the data reported in Sect. \ref{season}, 
 we also display the number of desk rejected papers by all editors during "vacation time" on Fig. \ref{Plot9rejdayvac}. 
 
 A similar  exponential-like behavior, as that found in    Fig. \ref{Plot2dayofweekrejectiony},  for the overall editor rejection day distribution, is found with a relaxation time decay  $\sim 2.545$.  Thus, one can consider that the editors are equally behaving during the academic year or during the vacation time.
 
 Notice that a conclusion based on the data examined in this appendix seems to indicate that editors work much at the beginning of the week.
 
 \vskip0.5cm
 \clearpage

\end{document}